\documentclass[5p,longtitle]{elsarticle}

\usepackage{lineno,hyperref}

\usepackage{color}
\usepackage{natbib}
\usepackage{amsmath}
\usepackage{booktabs}
\usepackage[normalem]{ulem}

\newcommand{\qsqU}{$(\text{GeV}/c)^2$}
\newcommand{\di}[0]{\mathrm{d}}
\newcommand{\xsecval}[5]{$#1 \ {}_{- \ #2}^{+ \ #3}\big\rvert_{\text{stat}} \ {}_{- \ #4}^{+ \ #5}\big\rvert_{\text{sys}}$}

\usepackage{mathtools}
\usepackage{siunitx}
\usepackage{tikz}

\usepackage{stackrel,amssymb}

\biboptions{numbers,sort&compress}

\usetikzlibrary{calc,arrows}
\usetikzlibrary{trees}
\usetikzlibrary{shapes}
\usetikzlibrary{positioning,shapes,shadows,arrows}
\usetikzlibrary{decorations.pathmorphing}
\usetikzlibrary{decorations.markings}
\usetikzlibrary{decorations.pathreplacing}
\usetikzlibrary{automata,positioning}
\tikzset{
    photon/.style={decorate, decoration={snake,pre length=2pt,post length=1pt, amplitude=1.5pt, segment length=4pt}, draw=red},
    photonarr/.style={draw=red, decoration={snake}, decorate, postaction={decoration={markings, mark=at position 1 with {\arrow[draw=red]{>}}}, decorate}},
    flash/.style={dashed, draw=red, postaction={decorate},
        decoration={markings,mark=at position 1 with {\arrow[draw=red]{>}}}},
	lepton/.style={draw=blue, postaction={decorate},
        decoration={markings,mark=at position .55 with {\arrow[draw=blue]{>}}}},
    nucleon/.style={draw=black, postaction={decorate},
        decoration={markings,mark=at position .55 with {\arrow[draw=black]{>}}}},
	nucleonsimple/.style={draw=black, double distance=4pt, postaction={decorate},
		decoration={markings,mark=at position .55 with {\arrow[draw=black, line width=0.3, scale=5]{>}}}},
    gluon/.style={decorate, draw=black,
        decoration={coil,amplitude=1.5pt, segment length=2pt}},
}

\newcommand{\xBj}[0]{x_{\text{Bj}}}
\newcommand{\LeftRight}{\leftrightarrows}

\DeclarePairedDelimiter\abs{\lvert}{\rvert}%
\DeclarePairedDelimiter\mean{\langle}{\rangle}%

\journal{Physics Letters B}

\begin{document}

\begin{frontmatter}

\title{Measurement of the 
cross section for hard exclusive $\pi^0$ leptoproduction\\
[15pt] {\normalsize COMPASS Collaboration}}

\author[turin_u]{M.~G.~Alexeev}
\author[dubna]{G.~D.~Alexeev}
\author[turin_u,turin_i]{A.~Amoroso}
\author[cern,illinois]{V.~Andrieux}
\author[dubna]{N.~V.~Anfimov}
\author[dubna]{V.~Anosov}
\author[dubna]{A.~Antoshkin}
\author[dubna,praguectu]{K.~Augsten}
\author[warsaw]{W.~Augustyniak}
\author[aveiro]{C.~D.~R.~Azevedo}
% ...
\address[aveiro]{University of Aveiro, I3N - Physics Department, 3810-193 Aveiro, Portugal}
% ...
\address[dubna]{Joint Institute for Nuclear Research, 141980 Dubna, Moscow region, Russia\fnref{E}}
% ...
\address[cern]{CERN, 1211 Geneva 23, Switzerland}
% ...
\address[praguectu]{Czech Technical University in Prague, 16636 Prague, Czech Republic\fnref{S}}
% ...
\address[triest_u]{University of Trieste, Dept.\ of Physics, 34127 Trieste, Italy}
\address[triest_i]{Trieste Section of INFN, 34127 Trieste, Italy}
\address[turin_u]{University of Turin, Dept.\ of Physics, 10125 Turin, Italy}
\address[turin_i]{Torino Section of INFN, 10125 Turin, Italy}
% ...
\address[illinois]{University of Illinois at Urbana-Champaign, Dept.\ of Physics, Urbana, IL 61801-3080, USA\fnref{Z}}
\address[warsaw]{National Centre for Nuclear Research, 00-681 Warsaw, Poland\fnref{a}}
% ...
\fntext[A]{Also at Instituto Superior T\'ecnico, Universidade de Lisboa, Lisbon, Portugal}
\fntext[B]{Also at Dept.\ of Physics, Pusan National University, Busan 609-735, Republic of Korea}
\fntext[B1]{Also at at Physics Dept., Brookhaven National Laboratory, Upton, NY 11973, USA}
\fntext[C]{Also at Abdus Salam ICTP, 34151 Trieste, Italy}
\fntext[D]{Supported by the DFG cluster of excellence `Origin and Structure of the Universe' (www.universe-cluster.de) (Germany)}
\fntext[F]{Supported by the DFG Research Training Group Programmes 1102 and 2044 (Germany)}
\fntext[G]{Also at Chubu University, Kasugai, Aichi 487-8501, Japan\fnref{V}}
\fntext[H]{Also at Dept.\ of Physics, National Central University, 300 Jhongda Road, Jhongli 32001, Taiwan}
\fntext[I]{Also at KEK, 1-1 Oho, Tsukuba, Ibaraki 305-0801, Japan}
\fntext[J]{Present address: Universit\"at Bonn, Physikalisches Institut, 53115 Bonn, Germany}
\fntext[K]{Also at Moscow Institute of Physics and Technology, Moscow Region, 141700, Russia}
\fntext[L]{Also at Yerevan Physics Institute, Alikhanian Br. Street, Yerevan, Armenia, 0036}
\fntext[M]{Also at Dept.\ of Physics, National Kaohsiung Normal University, Kaohsiung County 824, Taiwan}
\fntext[N]{Also at Institut f\"ur Theoretische Physik, Universit\"at T\"ubingen, 72076 T\"ubingen, Germany}
\fntext[O]{Also at University of Eastern Piedmont, 15100 Alessandria, Italy}
\fntext[P]{Present address: RWTH Aachen University, III.\ Physikalisches Institut, 52056 Aachen, Germany}
\fntext[Q]{Supported by BMBF - Bundesministerium f\"ur Bildung und Forschung (Germany)}
\fntext[R]{Supported by FP7, HadronPhysics3, Grant 283286 (European Union)}
\fntext[S]{Supported by MEYS, Grant LM20150581 (Czech Republic)}
\fntext[T]{Supported by B.Sen fund (India)}
\fntext[U]{Supported by FCT, COMPETE and QREN, Grants CERN/FP 116376/2010, 123600/2011 and CERN/FIS-NUC/0017/2015 (Portugal)}
\fntext[V]{Supported by MEXT and JSPS, Grants 18002006, 20540299, 18540281 and 26247032, the Daiko and Yamada Foundations (Japan)}
\fntext[W]{Supported by the Ministry of Science and Technology (Taiwan)}
\fntext[X]{Supported by the Israel Academy of Sciences and Humanities (Israel)}
\fntext[Y]{Supported by the Russian Federation  program ``Nauka'' (Contract No. 0.1764.GZB.2017) (Russia)}
\fntext[Z]{Supported by the National Science Foundation, Grant no. PHY-1506416 (USA)}
\fntext[a]{Supported by NCN, Grant 2017/26/M/ST2/00498 (Poland)}
\fntext[*]{Retired}
\fntext[**]{Deceased}
\cortext[cors]{Corresponding authors}



\sloppy
\begin{abstract}
{
  We report on a measurement of hard exclusive $\pi^0$ muoproduction on the proton by COMPASS
  using 160\,GeV/$c$ polarised $\mu^+$ and $\mu^-$ beams of the CERN SPS impinging on a liquid
  hydrogen target. From the average of the measured $\mu^+$ and $\mu^-$ cross sections, the
  virtual-photon proton cross section is determined as a function of the squared four-momentum
  transfer between initial and final proton in the range $0.08\,(\text{GeV/}c)^2 < |t| < 0.64\,(\text{GeV/}c)^2$.
  The average kinematics of the measurement are
  { $\langle Q^2 \rangle =2.0\; {(\text{GeV}/c)^2}$,
    $\langle \nu \rangle = 12.8\; {\text{GeV}}$, 
    $\langle x_{Bj} \rangle = 0.093  $ and 
    $\langle -t \rangle = 0.256\; {(\text{GeV}/c)^2}  $}.
  Fitting the azimuthal dependence reveals a combined contribution by transversely and longitudinally polarised photons of
  $(8.1 \ \pm \ 0.9_{\text{stat}}{}_{- \ 1.0}^{+ \ 1.1}\big\rvert_{\text{sys}})\,{\text{nb}}/{(\text{GeV}/c)^{2}}$,
  as well as transverse-transverse and longitudinal-transverse interference contributions of
  $(-6.0 \pm 1.3_{\text{stat}}{}_{- \ 0.7}^{+ \ 0.7}\big\rvert_{\text{sys}})\,{\text{nb}}/{(\text{GeV}/c)^{2}}$
  and $(1.4 \pm 0.5_{\text{stat}}{}_{- \ 0.2}^{+ \ 0.3}\big\rvert_{\text{sys}})\,{\text{nb}}/{(\text{GeV}/c)^{2}}$,
  respectively. Our results provide important input for modelling Generalised Parton Distributions.
  In the context of the phenomenological Goloskokov-Kroll model, the statistically significant
  transverse-transverse interference contribution constitutes clear experimental evidence
  for the chiral-odd GPD $\overline{E}_T$.
}
\end{abstract}



\begin{keyword}
Quantum chromodynamics, muoproduction, hard exclusive meson production, Generalised Parton Distributions, COMPASS
\end{keyword}

\end{frontmatter}


\section{Introduction}

Measurements of pseudoscalar mesons produced in hard exclusive lepton-nucleon scattering
provide
important data for phenomenological parameterisations of Generalised Parton Distributions (GPDs) \cite{Mueller:1998fv,Ji:1996ek, Ji:1996nm, Radyushkin:1996nd, Radyushkin:1997ki}.
In the past two decades, GPDs have shown to be a very rich 
and useful construct for both experiment 
and theory as their determination allows for a detailed description of the parton structure of the nucleon. In particular, 
GPDs correlate transverse spatial positions and longitudinal momentum fractions 
of the partons in the nucleon. {They embed parton distribution functions and nucleon form factors, and they give access to energy-momentum-tensor form factors}.
For each quark flavour, there exist four parton-helicity-conserving (chiral-even) GPDs, denoted 
$H$, $\tilde{H}$, $E$, and $\tilde{E}$, 
and four parton-helicity-flip (chiral-odd) 
GPDs, denoted $H_T$, $\tilde{H}_T$, $E_T$, and $\tilde{E}_T$.
While hard production of vector mesons is sensitive primarily to 
the GPDs $H$ and $E$, 
the production of pseudoscalar mesons by longitudinally polarised virtual photons is sensitive to $\tilde{H}$ and $\tilde{E}$ in the leading-twist description.

Contributions from transversely polarised virtual photons to the production of spin-0
mesons are expected to be suppressed in the production amplitude by $1/Q$~\cite{Collins-1996-ID17},
where $Q^2$ is the virtuality of the photon $\gamma^*$ that is exchanged between muon and proton. However, 
experimental data on exclusive $\pi^+$ production from HERMES~\cite{hermes1} and on exclusive $\pi^0$
production from JLab CLAS~\cite{clas3, clas2, clas1, clas4} and Hall A~\cite{hallA1, hallA2, hallA3} 
suggest that such contributions are substantial.
In the GPD formalism such contributions are possible
if a quark helicity-flip GPD couples to a twist-3 wave function~\cite{GK2011, Ahmad-2008-ID20}.
In the framework of the phenomenological model of Ref.~\cite{GK2011},
pseudoscalar-meson production is described by the GPDs
$\tilde{H}$, $\tilde{E}$, $H_T$ and $\overline{E}_{T} = 2\tilde{H}_T + E_T$.
Different sensitivities to these GPDs are expected when comparing
$\pi^+$ vs. $\pi^0$ production.
When taking into account the relative signs and sizes of these GPDs for $u$ and $d$ quarks, 
the different quark flavour contents of these mesons lead to different predictions for the $\abs{t}$-dependence of 
the cross section, especially at small values of $|t|$.
Here, $t$ is the square of the four-momentum transfer between initial and final nucleon. 
The production of $\pi^+$ mesons
is dominated by the contributions 
from longitudinally polarised virtual photons, of which a major part 
originates from pion-pole exchange 
that is the main contributor to $\tilde{E}$. 
Also the contributions 
from $\tilde{H}$ and $H_T$ are significant, and 
there is a strong cancellation between the contributions from  
$\overline{E}_{T}$ for $u$ and $d$ quarks. 
On the contrary, in the case of $\pi^0$ production
there is no pion-pole exchange,
the contributions from $\tilde{H}$ and $H_T$ are small and a large contribution from
transversely polarised photons is generated mainly by $\overline{E}_{T}$.

These differences between $\pi^+$ and $\pi^0$ production 
are experimentally supported. While for $\pi^+$ 
production a fast decrease of the cross section with increasing $|t|$ is predicted by theoretical
models and confirmed by the experimental results from HERMES~\cite{hermes1}, 
for $\pi^0$ production a dip is expected as $|t| \rightarrow 0$~\cite{GK2011} and confirmed 
by results in the JLab kinematic domain~\cite{clas2,clas1,hallA2}.
Constraints for modelling the poorly known GPD $\overline{E}_{T}$ 
were obtained in a lattice-QCD study~\cite{QCDSF} of its moments. The 
COMPASS results on exclusive $\pi^0$ production 
in muon-proton scattering {presented in this Letter} provide 
new input for modelling 
this GPD and chiral-odd (`transversity') GPDs in general.

\section{Formalism}
\begin{figure}[h!] 
\begin{center}  
  \includegraphics[width=0.8\linewidth]{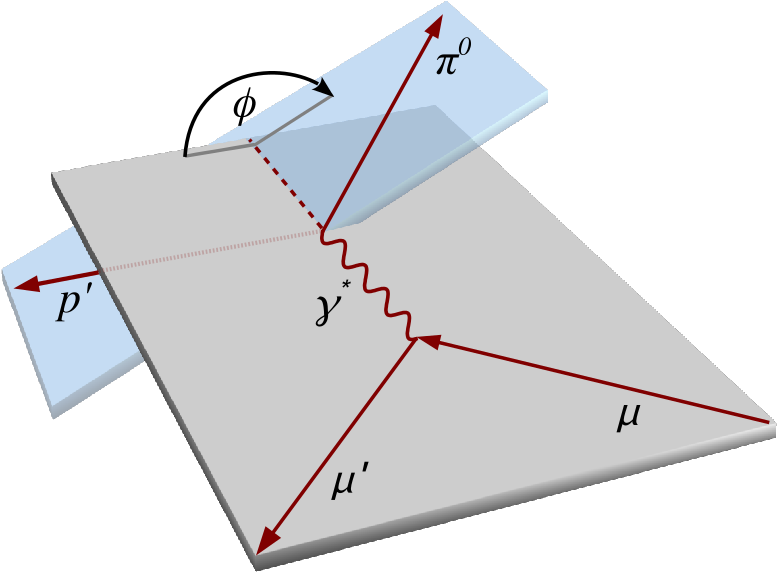}
\end{center} 
\vspace{-0.25cm}
  \caption{\label{fig:phi_angle}
  Definition of $\phi$, the azimuthal angle between the lepton-scattering and $\pi^0$-production planes.
  }
\end{figure}

The reduced cross section for 
hard exclusive meson production by scattering 
a polarised lepton beam
off an unpolarised proton target reads:
\begin{align}
\label{eq:po_x_sec}
&\frac{\di^2 \sigma _{\gamma^* p }^{\LeftRight}}{ \di t \di \phi}  = \frac{1}{2\pi}
\Big[\frac{\di \sigma_{T}}{\di t} + \epsilon \frac{\di \sigma_{L}}{\di t}
+ \epsilon \cos \left( 2\phi \right) \frac{\di \sigma_{TT}}{\di t}\\
\nonumber
&+ \sqrt{2 \epsilon \left( 1 + \epsilon \right)} \cos \phi 
\frac{\di \sigma_{LT}}{\di t}
\mp |P_l| \sqrt{2 \epsilon(1-\epsilon)} \sin{\phi} \frac{\di \sigma_{LT}'}{\di t}\Big], &
\end{align}
where the sign $\mp$ of the lepton beam polarisation $P_l$ corresponds to negative and positive helicity of the incoming lepton, respectively, denoted by $\LeftRight$.
The conversion from the lepton-nucleon cross section to the virtual-photon nucleon cross
section, using the one-photon-exchange approximation, is explained in Sect.~\ref{sec:x_sec}.
The contribution to the cross section from transversely
(longitudinally) polarised virtual photons is denoted by $\sigma_T$ ($\sigma_L$). 
The symbols $\sigma_{TT}$ and $\sigma_{LT}, \sigma'_{LT}$ denote contributions from the interference between transversely and longitudinally polarised virtual photons, respectively, with transversely polarised ones
The factor
\begin{equation}
\epsilon=\frac{1-y-\frac{y^2\gamma^2}{4}}{1-y+\frac{y^2}{2}+\frac{y^2\gamma^2}{4}}
\end{equation}
is the virtual-photon polarisation parameter and $\phi$ 
is the azimuthal angle between the 
lepton scattering plane and the hadron production plane, see Fig.~\ref{fig:phi_angle}. 
Here, $Q^2=-(k_{\mu}-k_{\mu'})^2$ is the photon virtuality,
$\nu=(k_{\mu}^0-k_{\mu'}^0)$ the energy of the virtual photon in the target rest frame, $y=\nu/k_{\mu}^0$ 
and $\gamma^2=Q^2/\nu^2$, where $k_{\mu}$ and
$k_{\mu'}$ denote the four-momenta of the incoming and the scattered muon in the laboratory system, respectively. 

The spin-independent cross section can be obtained by averaging the two spin-dependent cross sections,
\begin{equation}
\label{eq:sum_x_sec}
\frac{\di^2 \sigma _{\gamma^* p }}{ \di t \di \phi}  = \frac{1}{2}\Bigl(\frac{\di^2 \sigma _{\gamma^* p } ^{\leftarrow} }{ \di t \di \phi}+\frac{\di^2 \sigma _{\gamma^* p }  ^{\rightarrow}}{ \di t \di \phi}\Bigr). 
\end{equation}
When forming this average, the last term in Eq.~(\ref{eq:po_x_sec}) 
cancels if the magnitude $|P_l|$ of the beam polarisation is the same for
measurements with $\mu^+$ and $\mu^-$ beam, so that
\begin{align}
\label{eq:new_unpo}
\frac{\di^2 \sigma _{\gamma^* p }}{ \di t \di \phi} = 
&\,\frac{1}{2\pi}
\Big[\frac{\di \sigma_{T}}{\di t} + \epsilon \frac{\di \sigma_{L}}{\di t}
+ \epsilon \cos \left( 2\phi \right) \frac{\di \sigma_{TT}}{\di t}\\
\nonumber
&+ \sqrt{2 \epsilon \left( 1 + \epsilon \right)} \cos \left( \phi \right) 
\frac{\di \sigma_{LT}}{\di t}\Big]. &
\end{align}

The individual contributions appearing in Eq.~(\ref{eq:new_unpo}) are related to convolutions 
of GPDs and meson distribution amplitudes with 
individual hard scattering amplitudes~\cite{GK2011, clas1}:
\begin{align}
\label{equ::relations_s}
    \frac{\di \sigma_{T}}{\di t} &\propto \Big[ (1-\xi^2) \abs{\mean{H_T}}^2 - \frac{t'}{8M^2}\abs{\mean{\overline{E}_{T}}}^2 \Big],\\
    \frac{\di \sigma_{L}}{\di t} &\propto 
    \Big[
        (1-\xi^2) \abs{\mean{\tilde{H}}}^2 \nonumber \\
        & - 2\xi^2 Re\left[ \mean{\tilde{H}}^\ast\mean{\tilde{E}} \right]
        - \frac{t'}{4M^2}\xi^2\abs{\mean{\tilde{E}}}^2
    \Big],\\
    \frac{\di \sigma_{TT}}{\di t} &\propto t' \abs{\mean{\overline{E}_T}}^2,\\ \label{equ::relations_e}
    \frac{\di \sigma_{LT}}{\di t} &\propto \xi \sqrt{1-\xi^2}\sqrt{-t'} Re\left[ \mean{H_T}^\ast\mean{\tilde{E}} \right].
\end{align}
Here, the aforementioned convolutions are denoted by triangular brackets,
$t' = t - t_{min}$ with $|t_{min}|$ being the kinematically smallest possible 
value of $|t|$, and $M$ is the mass of the proton.
The quantity $\xi$ is equal to one half of the longitudinal momentum fraction 
transferred between the initial and final proton and can be approximated at COMPASS kinematics as
\begin{equation}
\xi \approx \frac{\xBj}{2 - \xBj},
\end{equation}
where $\xBj=Q^2/(2M \nu)$.

\section{Experimental set-up and data selection}
\label{sec:ev_sel}
The main component of the COMPASS set-up 
is the two-stage magnetic spectrometer. Each spectrometer
stage comprises a dipole magnet complemented by a variety of tracking detectors, a muon filter
for muon identification and an electromagnetic (ECal) as well as a hadron calorimeter.
A detailed description of the set-up can be found in Refs.~\cite{spec1,spec2,spec3}.

The data used for this analysis were collected within four weeks in 2012, during which the COMPASS spec\-tro\-meter was complemented by a 2.5\,m long liquid-hydrogen target surrounded
by a time-of-flight (TOF) system, and a third electromagnetic calorimeter
that was placed directly downstream of the target. The 
TOF detector consisted of 
two cylinders, each made of 24 scintillating-counter slats, 
mounted concentrically around the target.

In order to determine the spin-independent cross section through Eq.~(\ref{eq:sum_x_sec}), data with $\mu^+$ and $\mu^-$ beam 
were taken separately. The natural polarisation of the muon beam 
provided by the CERN SPS
originates from the parity-violating decay in flight of the parent meson, 
which implies opposite polarisation for $\mu^+$ and $\mu^-$ beams.
Within regular time intervals 
during the measurement, charge and polarisation of the muon beam were swapped simultaneously.
In total, a luminosity of 18.9\,pb$^{-1}$ was collected for 
the $\mu^+$ beam with negative polarisation and  23.5\,pb$^{-1}$ 
for the $\mu^-$ beam with positive polarisation. 
For both beams, the absolute value of the average beam 
polarisation is about 0.8 with an uncertainty of about 0.04.

In the data analysis, $\pi^0$ mesons
are selected by their dominant two-photon decay.
At least two neutral clusters are required that had to be 
detected above the respective threshold in one of the electromagnetic calorimeters, 
in conjunction with an interaction vertex re\-con\-structed 
within the target using the incoming and outgoing muon tracks. 
The outgoing muon is identified by 
requiring that it has the same charge as the beam particle and 
traverses more than 15 radiation lengths. As neutral cluster 
we denote a reconstructed calorimeter cluster 
that is not associated to a charged track, 
thereby including any cluster in case of the most upstream calorimeter
that had no tracking system in front. 

\begin{figure}[h!]
\includegraphics[width=0.49\textwidth]{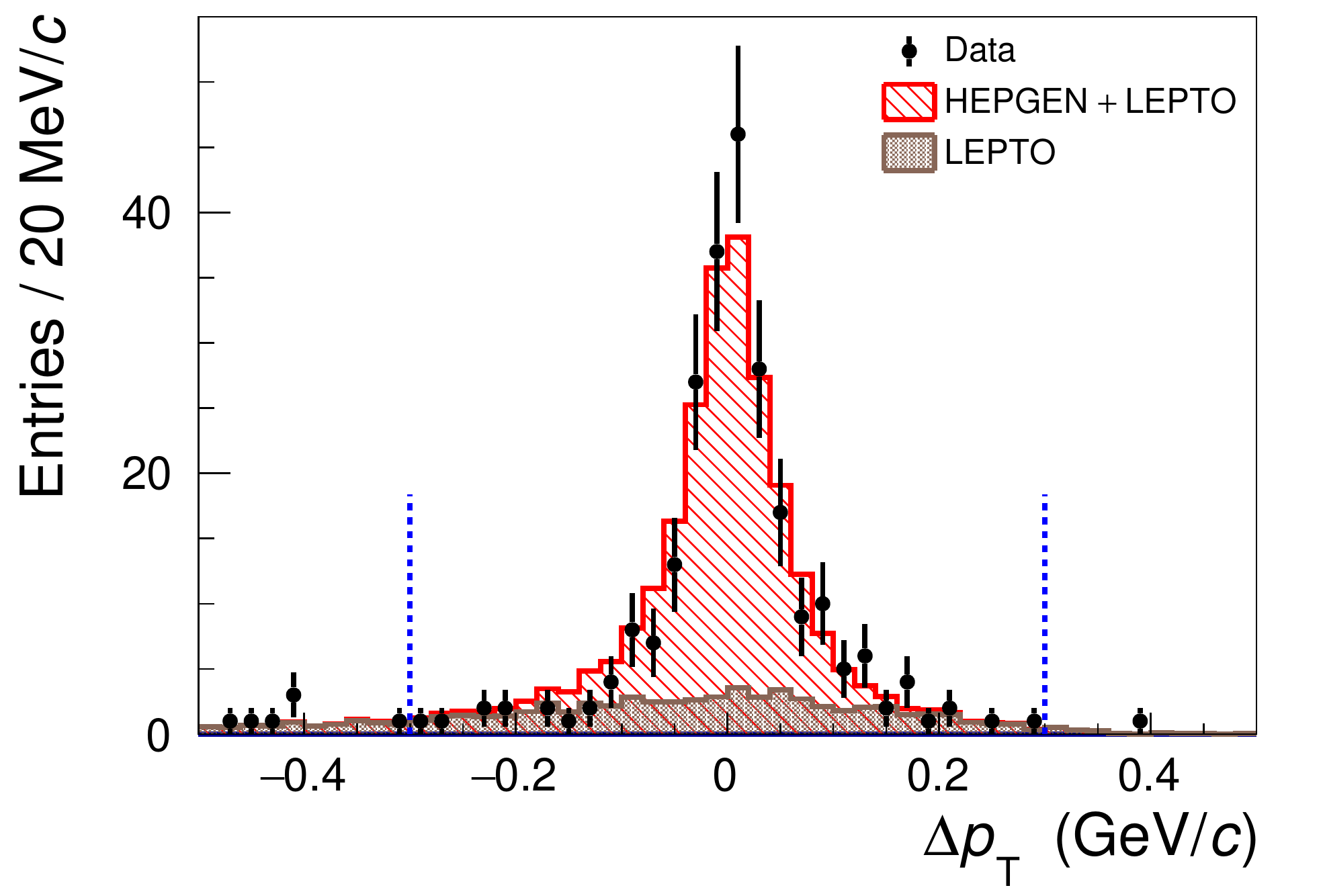}
\caption
{\label{fig:ex_vars}
Measured and simulated distribution of the
difference between predicted and reconstructed 
transverse momentum, $\Delta p_{\text{T}}$, of the recoil 
proton for the kinematic region described in the text. 
The vertical lines indicate the 
constraints applied for the selection of events. 
The quantity $p_{\text{T}}$ is defined in the laboratory system.
Error bars denote statistical uncertainties.}
\end{figure}

For each interaction vertex and each combination of two neutral clusters, the 
kinematics of the recoil
proton are predicted from the four-momentum balance of the 
ana\-ly\-sed process, $\mu p \rightarrow \mu' p' \pi^0$, $\pi^0 \rightarrow \gamma \gamma$, 
by using the reconstructed spectrometer information, i.e.\ the
vertex position, the momenta of the incoming and outgoing muons 
as well as the energy and position of the two clusters.
The predicted properties of the recoil proton $p'$ 
are compared to the properties of each track candidate as 
reconstructed by the TOF system. Note that the four-momentum of the recoil
proton is determined by the target TOF system based on 
the assumption that the reconstructed track belongs to a proton.
Figure~\ref{fig:ex_vars} shows an 
example for the result of 
such a comparison, including also
the corresponding constraints applied for the selection of events. 
The Monte Carlo yields shown in this figure, denoted as HEPGEN and LEPTO, will be explained in more detail in Sect.~\ref{sec:bkg}.

\begin{figure}[h!]
\begin{center}  
  \includegraphics[width=\linewidth]{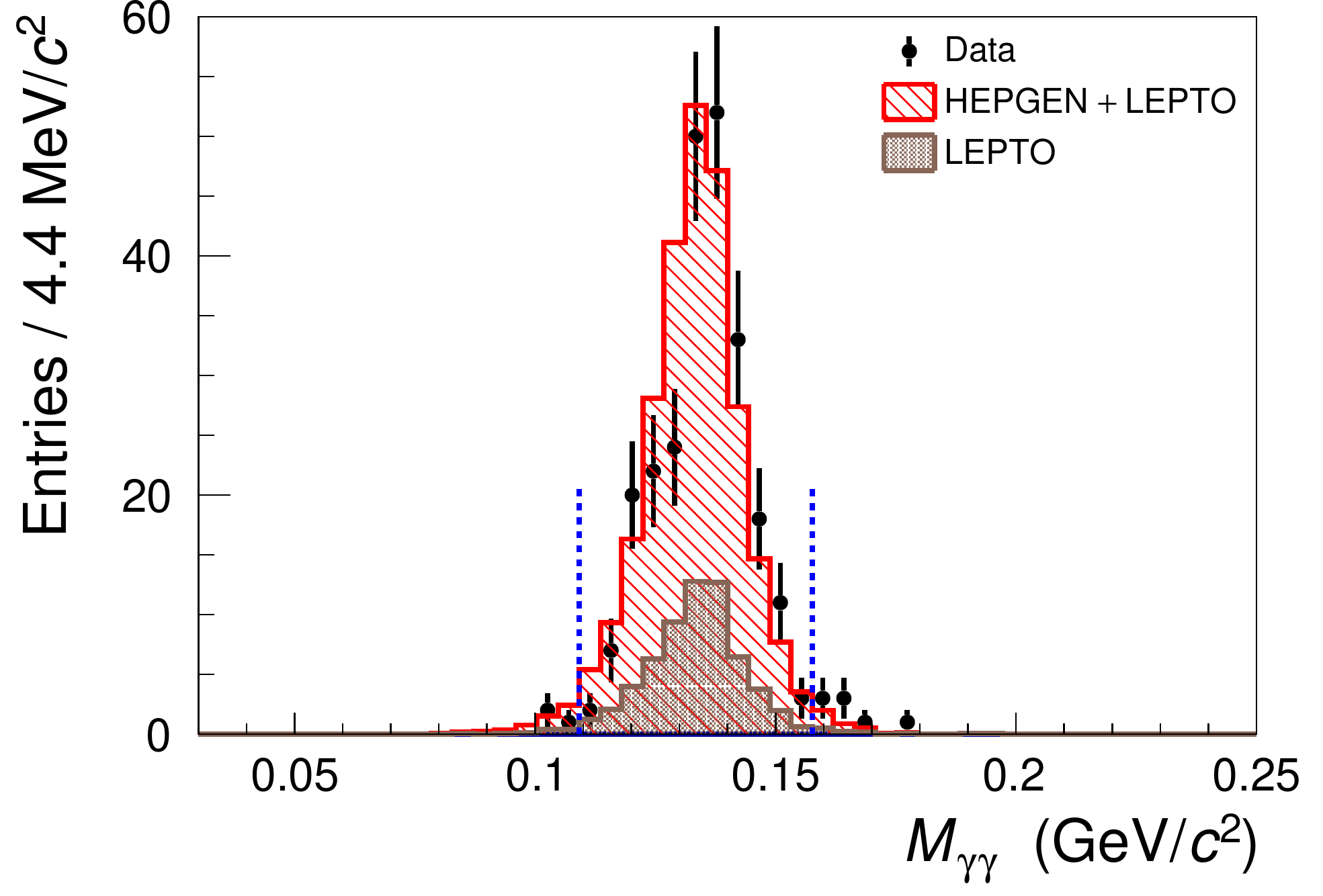}
\end{center} 
\vspace{-0.25cm}
  \caption{\label{fig:pi0_mass}Distribution of the invariant mass
  $M_{\gamma \gamma}$
  of the two-photon system. 
  Otherwise as in figure~\ref{fig:ex_vars}.
  }
\end{figure}

In the case that 
more than one combination of vertex, cluster pair and recoil-track candidate 
exist that satisfy the aforementioned selection
criteria for a given event, this 
event is excluded from the analysis. 
Figure~\ref{fig:pi0_mass} shows the 
two-photon mass distribution in 
the region close to the nominal $\pi^0$ mass  
together with the constraints applied to select $\pi^0$ mesons.

In order to further enhance the purity of the 
selected data and to improve the precision of the particle 
kinematics at the interaction vertex, a kinematic fit for the 
exclusive reaction $\mu p \rightarrow \mu' p' \pi^0$ 
is performed, which requires a single $\pi^0$ to decay into the two photons selected as described above.

With the selection procedure described above, data are analysed
in the kinematic range 
\begin{eqnarray*}
0.08\,(\text{GeV}/c)^2 & < \, |t| \,  < & 0.64\,(\text{GeV}/c)^2, \\
1\,(\text{GeV}/c)^2    & < \, Q^2 \,  < & 5\,(\text{GeV}/c)^2     \\ 
\text{and \hspace{0.8cm} }8.5\,\text{GeV} & < \, \nu \,  < & 28\,\text{GeV} \text{.} 
\end{eqnarray*}
In addition, two 
reference samples are selected in a wider kinematic range, which are denoted as signal and 
background sample. Apart from the extended kinematic range given below, the signal sample corresponds to the aforementioned selections.
In contrast to the signal sample, the background sample contains only events with more than one combination of vertex, cluster pair and recoil-track candidate. 
Apart from the small peak at zero, it contains non-exclusive events. The purpose of the 
reference samples is explained in the following section.

\section{Estimation of the background contribution}
\label{sec:bkg}

The main background to exclusive $\pi^0$ muoproduction 
originates from non-exclusive deep-inelastic scattering processes. 
In such processes, low-energy hadrons are produced in addition to 
the $\pi^0$, which remain undetected in the apparatus.
In order to estimate the background contribution, two Monte Carlo generators are employed.

First, the
LEPTO 6.5.1 generator with the high-$p_{T}$ COMPASS tuning~\cite{highpt} is used to describe the non-exclusive fraction of events.
Secondly, the HEPGEN++ $\pi^0$ generator is used to 
model the kinematics of single $\pi^0$ muoproduction~\cite{HEPGEN,HEPGEN_pp}, in the following denoted
by HEPGEN. The generated events from both generators are
independently passed through a complete 
description of the COMPASS set-up~\cite{TGeant},
and the resulting simulated data are 
treated in the same way as it is done for real data.

\begin{figure}[h!]
\begin{center}
\begin{tabular}{c}
\includegraphics[width=0.49\textwidth]{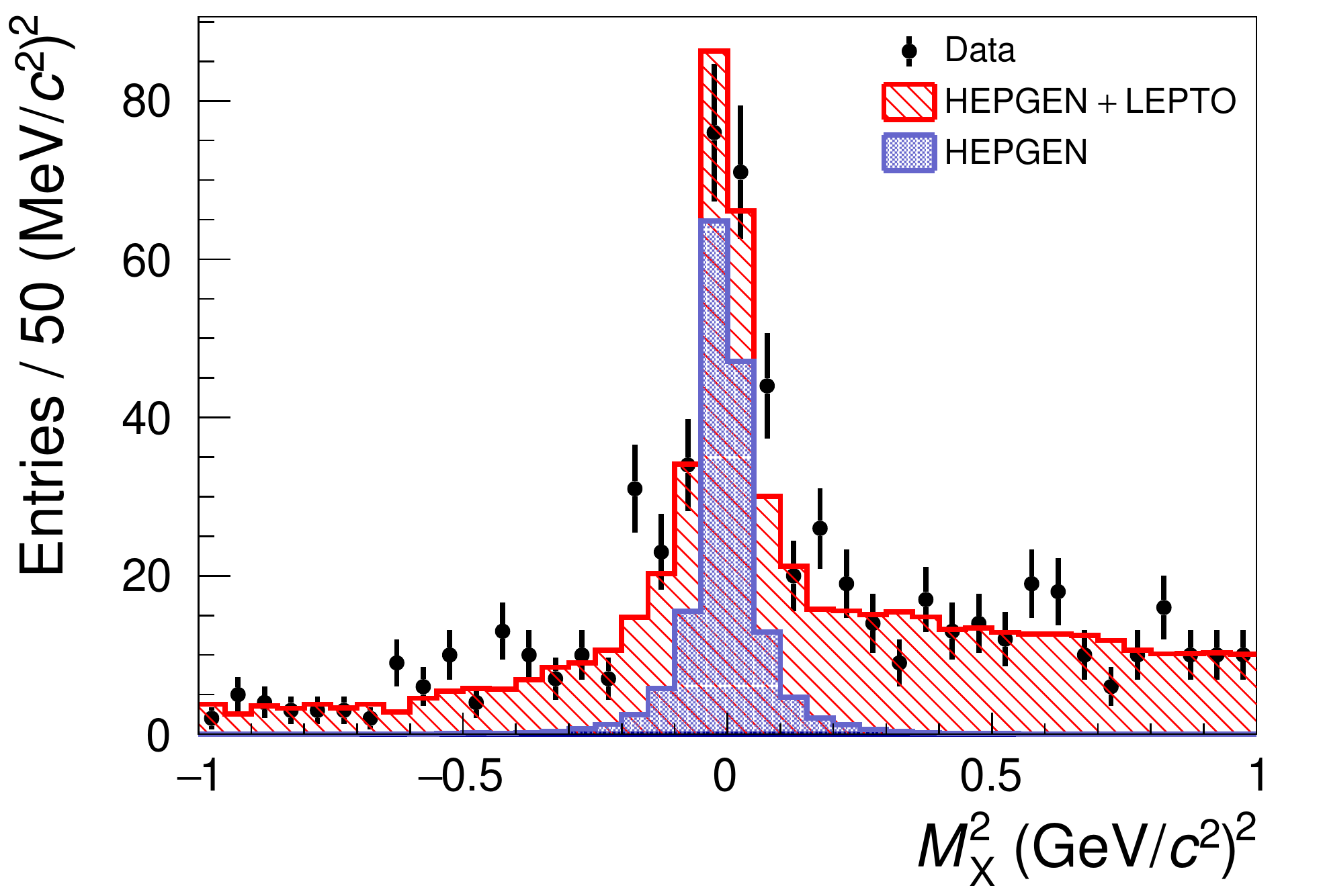}
\\
\includegraphics[width=0.49\textwidth]{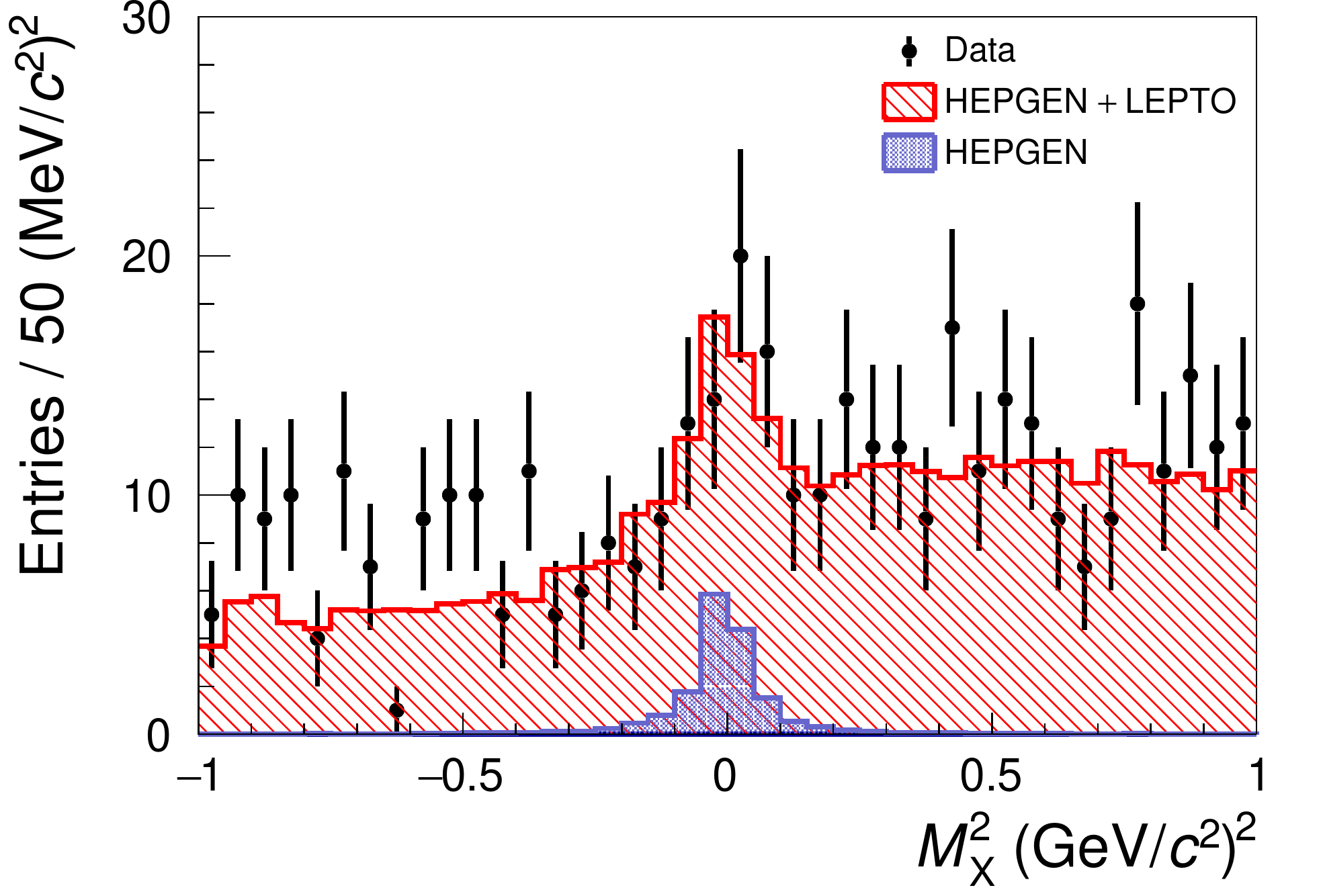}
\end{tabular}
\end{center}
\vspace{-0.5cm}
\label{fig:bkg} 
\caption{Distributions of the undetected mass $M_X^2$ for the 
signal (top) and background (bottom) reference
samples, which are selected as described in Sect.~\ref{sec:ev_sel}
in the extended kinematic range 
$Q^2>1(\text{GeV}/c)^2$, $y>0.05$ and $|t_{\text{TOF}}|>0.08(\text{GeV}/c)^2$. 
Here, the quantities $y$ and $t_{\text{TOF}}$ denote the fractional energy loss of the muon and
the squared four-momentum transfer to the recoil
proton as measured by the target TOF system, respectively. 
Simulated data are also shown (see text).
Note that events with the topology of exclusive $\pi^0$ production
were removed from the LEPTO sample. Error bars denote statistical uncertainties.}
\end{figure}

As there exists essentially no information on the cross section of exclusive
$\pi^0$ production in the kinematic domain of COMPASS, 
the two reference samples 
described in Sect.~\ref{sec:ev_sel} 
are used to normalise the HEPGEN and LEPTO Monte Carlo yields.
Using several variables, the kinematic information from 
beam and spectrometer measurements as well as that of 
the recoil-proton candidates are compared between experimental data and the
two simulations in order to determine the best normalisation of each 
simulated data set relative to that of the experimental data. 
As an example of such a comparison, the undetected mass
\begin{equation}
M_{\text{X}}^2=(k_{\mu}+p_{p}-k_{\mu'}-p_{p'}-p_{\gamma_1}-p_{\gamma_2})^2
\end{equation}
is shown in Fig.~\ref{fig:bkg}.
Here, the four-momenta are denoted by
$p_{p}$ and $p_{p'}$ for the target and recoil proton, respectively, 
and by $p_{\gamma_1}$ and $p_{\gamma_2}$ for the two produced 
photons. In addition to the measured data points, the HEPGEN 
simulation and the sum of the HEPGEN and LEPTO 
simulations are shown. 
In order to estimate the amount of non-exclusive background, the 
simulated data are scaled such that they describe the data for both
reference samples. The scaling factor for the LEPTO Monte Carlo yield, which is denoted by $f^\pm$, 
will be used in Sect.~\ref{sec:x_sec} to normalise this simulation when correcting the data for background.

The resulting fraction of non-exclusive background in the 
data is estimated to be 
{$(29 {}_{- \ 6}^{+ \ 2}\big\rvert_{\text{sys}}    )\% $}. 
Here, the uncertainty is estimated by comparing the scaling factors
extracted for various variables and by using several extraction methods for
the scaling factors. Details are given in Ref.~\cite{matze}.
Contributions of other background sources
are found to be negligible. 
For example, the production of single $\omega$ mesons, where the $\omega$ decays into a $\pi^0$ and a photon 
that remains undetected, was found in Monte Carlo studies to contribute at the level of 1\%~\cite{matze}.

\section{Determination of the cross section}
\label{sec:x_sec}
The virtual-photon proton cross section is obtained from the measured
muon-proton cross section using
\begin{equation}
\frac{\di^2 \sigma}{\di |t| \di \phi } =
\frac{1}{\Gamma(Q^2,\nu,E_{\mu})}\,\frac{\di \sigma^{\mu p}}{\di Q^2
  \di \nu\di \phi \di |t|},
\end{equation}
where the transverse virtual-photon flux is given by 
\begin{equation}  
\label{eq:photon_flux}
\begin{split}
\Gamma(Q^2,\nu,E_{\mu}) & = \frac{\alpha_{\text{em}}  (1- \xBj)}{2 \pi Q^2 y E_{\mu}} \Bigg[ y^2 \bigg(1 - \frac{2 m_{\mu}^2}{Q^2} \bigg) \\
& + \frac{2}{1+Q^2/\nu^2} \bigg(1-y - \frac{Q^2}{4E_{\mu}^2} \bigg) \Bigg].
\end{split}
\end{equation}
Here, $\alpha_{\text{em}}$ denotes the electromagnetic fine structure constant and $E_\mu$ the zero-th component of $k_{\mu}$.

For the cross section determination, the HEPGEN Mon\-te Carlo simulation described in Sect.~\ref{sec:bkg} is used. The acceptance $a(\Delta \Omega_{klmn})$ is calculated in a four-dimensio\-nal grid 
as the number of reconstructed events divided by the number of generated events 
using 8 bins in $\phi$, 5 in $|t|$, 4 in $Q^2$ and 4 in $\nu$.
The phase-space element is given by $\Delta \Omega_{klmn}=\Delta \phi_k \Delta |t|_l \Delta Q^2_m \Delta \nu_n$.
The spacing of the grid is given in Table~\ref{tab:1}.
\begin{table}[h!] 
\caption{\label{tab:1}
Four-dimensional grid used for the calculation of the acceptance. The full width of the respective dimension is given 
in the bottom row of the table.}
\footnotesize
\begin{center}
\begin{tabular}{l l l l}
$\phi$ /rad                                            & $|t|$ /\qsqU  &$Q^2$ /\qsqU  & $\nu$ /GeV        \\[2pt]
\hline \\[-6pt]
$-\pi$ \ -- $\frac{-3\pi}{4}$                           & 0.08 -- 0.15 & 1\textcolor{white}{.00} -- 1.5   & \textcolor{white}{0}8.5\textcolor{white}{0} -- 11.45 \\[2pt]
$\frac{-3\pi}{4}$ -- $\frac{-\pi}{2}$                   & 0.15 -- 0.22 & 1.5\textcolor{white}{0} -- 2.24 & 11.45 -- 15.43   \\[2pt]
\multicolumn{1}{c}{.}                                   & 0.22 -- 0.36 & 2.24 -- 3.34                    & 15.43 -- 0.78    \\[2pt]
\multicolumn{1}{c}{.}                                   & 0.36 -- 0.5  & 3.34 -- 5                       & 20.78 -- 28      \\[2pt]
\multicolumn{1}{c}{.}                                   & 0.5\textcolor{white}{0} -- 0.64 &              &                  \\[2pt]
$\frac{3\pi}{4}$ \ \ -- $\textcolor{white}{-} \pi$      &              &                                 &                  \\[2pt]
\hline \\[-8pt]
${\Delta} \phi$/rad                                     & ${\Delta}|t|$/\qsqU  &${\Delta} Q^2$/\qsqU  & ${\Delta} \nu$/GeV \\[2pt]
\hline \\[-6pt]
$2\pi$                                                         & 0.56 & 4     & 19.5     \\[2pt]
\end{tabular}
\end{center}
\end{table}

In each four-dimensional bin, the experimental yield corrected for background according to the LEPTO
simulations is obtained as
\begin{equation}
\label{eq:yield}
\begin{split}
\mathcal{Y}^{\pm}_{klmn} & =
\sum_{i=1}^{N_{\text{data}}^{\pm,\Delta \Omega_{klmn}}}\hspace{-5pt}\frac{1}{\Gamma(Q^2_i,\nu_i,E_{\mu, i})} \\
  & - f^{\pm} \sum_{i=1}^{N_{\text{L}}^{\pm,\Delta \Omega_{klmn}}}\hspace{-5pt}\frac{1}{\Gamma(Q^2_i,\nu_i,E_{\mu, i})}.
\end{split}
\end{equation}
Here, $N_{\text{data}}^{\pm,\Delta \Omega_{klmn}}$ is the number of measured
events and $N_{\text{L}}^{\pm,\Delta \Omega_{klmn}}$ the number of
LEPTO events within the phase-space element $\Delta
\Omega_{klmn}$. The second sum represents the LEPTO simulations that
are appropriately normalised by the factor $f^{\pm}$, which 
was introduced in Sect.~\ref{sec:bkg}.
Each event is weighted with the transverse virtual-photon flux $\Gamma(Q^2_i,\nu_i,E_{\mu, i})$ to obtain
the virtual-photon yield from the measured yields for muon-proton
interactions.

The spin-dependent virtual-photon proton cross sections measured with positively or negatively charged muons are determined in each of the
($\phi_k$, $|t|_l$) bins as luminosity-normalised experimental
yield averaged over
the measured ranges
$\Delta Q^2=4$\,\qsqU and $\Delta \nu = 19.5$\,GeV as
\begin{equation}
\label{eq:extract}
\Big\langle\frac{\di^2 \sigma}{\di |t| \di \phi}\Big\rangle_{\Delta \Omega_{kl}}^{\pm} =
\frac{1}{\mathcal{L^{\pm}} \Delta \Omega_{kl}}
\sum_{mn} \frac{\mathcal{Y}^{\pm}_{klmn}}{a(\Delta \Omega_{klmn})}.
\end{equation}
Here, 
$\Delta \Omega_{kl}= \Delta \phi_k \Delta |t|_l \Delta Q^2 \Delta \nu$, $\mathcal{L}^{\pm}$ denotes the luminosity and
$a(\Delta \Omega_{klmn})$ the acceptance in the
phase-space element $\Delta \Omega_{klmn}$.

The spin-independent virtual-photon proton cross section is obtained according to Eq.~(\ref{eq:sum_x_sec}) as average of the two spin-dependent cross sections given in Eq.~(\ref{eq:extract}):
\begin{equation}
\Big\langle\frac{\di^2 \sigma}{\di |t| \di \phi} \Big\rangle_{\Delta \Omega_{kl}}=
\frac{1}{2}\Biggl(\Big\langle\frac{\di^2 \sigma}{\di |t| \di \phi} \Big\rangle_{\Delta \Omega_{kl}}^{+}+
\Big\langle\frac{\di^2 \sigma}{\di |t| \di \phi} \Big\rangle_{\Delta \Omega_{kl}}^{-}\Biggr).
\end{equation}
The cross section integrated over the full $2\pi$-range in $\phi$ is obtained as
\begin{equation}
\label{eq:t}
\Big\langle\frac{\di \sigma}{\di |t|} \Big\rangle_{\Delta \Omega_{l}} =
\sum_{k} \Delta \phi_k \Big\langle\frac{\di^2 \sigma}{\di |t| \di \phi} \Big\rangle_{\Delta \Omega_{kl}},
\end{equation}
with $\Delta \Omega_l=\Delta |t|_l \Delta Q^2 \Delta \nu$.
Similarly, the $|t|$-averaged cross section in the measured range is given by
\begin{equation}
\label{eq:ohi}
\Big\langle\frac{\di^2 \sigma}{\di |t| \di \phi} \Big\rangle_{\Delta \Omega_{k}} =
\frac{1}{\Delta |t|}
\sum_{l} \Delta |t|_l \Big\langle\frac{\di^2 \sigma}{\di |t| \di \phi} \Big\rangle_{\Delta \Omega_{kl}},
\end{equation}
with $\Delta \Omega_k= \Delta \phi_k \Delta |t| \Delta Q^2 \Delta \nu$.

The systematic uncertainties on the extracted values of the cross section
are shown in Table~\ref{tab:sys_t}, arranged in three groups.
The first group contains the systematic uncertainties on the determination of the beam flux,
possible systematic effects related to the uncertainty on the energy thresholds for the
detection of the low-energetic photon in the electromagnetic calorimeters, and the 
uncertainty on the determination of the acceptance. The second group contains the systematic
uncertainties related to a variation of the energy and momentum balance of the kinematic fit,
the influence of background originating from the production of $\omega$ mesons and the estimated
influence of radiative corrections~\cite{rad_cor,matze}. The largest systematic effects appear in the third group,
which contains the uncertainty related to the estimation of non-exclusive background as described
in Sect.~\ref{sec:bkg}, and that related to an observed mismatch between the measured single-photon
yield in the 2012 COMPASS data and a corresponding Monte Carlo simulation of the Bethe-Heitler process.
The latter effect was observed in Refs.~\cite{philipp,dvcs_paper} in a kinematic region where
single-photon production is dominated by the Bethe-Heitler cross section, which is calculable at the
percent level. The total systematic uncertainty $\Sigma$ is obtained by quadratic summation of its components
for each bin separately.
\begin{table}[h!]
\centering
\caption[Syst. uncertainty in $t$, summary]{Summary of the estimated relative systematic uncertainties for the $|t|$ and $\phi$-dependent cross sections and the integrated cross section. The values are given in percent. Note that the uni-directional uncertainty
$\sigma_{\uparrow}$ ($\sigma_{\downarrow}$) has to be used with positive (negative) sign.}
\label{tab:sys_t}
\small
\begin{tabular}{  c   r    r    r    r   r  r}
  source &  $\sigma^t_{\uparrow}$ & $\sigma^t_{\downarrow}$ & $\sigma^{\phi}_{\uparrow}$ & $\sigma^{\phi}_{\downarrow}$&$\sigma_{\uparrow}$&$\sigma_{\downarrow}$\\[2pt]
\toprule
 $\mu^+$ flux  &        2& 2& 2& 2& 2& 2\\[1pt]
 $\mu^-$ flux  &        2& 2& 2& 2& 2& 2\\[1pt]
 ECal threshold &       5& 5& 5& 5& 5& 5\\[1pt]
 acceptance &           4& 7& 4& 7& 4& 7\\[2pt]
 \hline\\[-6pt]
 kinem. fit &           0& 7& 0& 7& 0& 7\\[1pt]
$\omega$ background &   0& 1& 0& 1& 0& 1 \\[1pt]
 rad. corr. &           2& 5& 2& 5& 2& 5\\[2pt]
 \hline\\[-6pt]
 $\mu^+$ event loss &   4--13&    0&0--12&0--5 & 9& 0\\[1pt]
 \textsc{Lepto} norm. & 5--28&3--11&5--51&3--21& 8& 3\\[2pt]
\bottomrule\\[-6pt]
 $\sum$ &              12--29&10--14&12--53&12--24&14& 13\\
\end{tabular}
\end{table}

\section{Results}
\label{sec:results}
\begin{figure}[h!]
\begin{center}  
  \includegraphics[width=\linewidth]{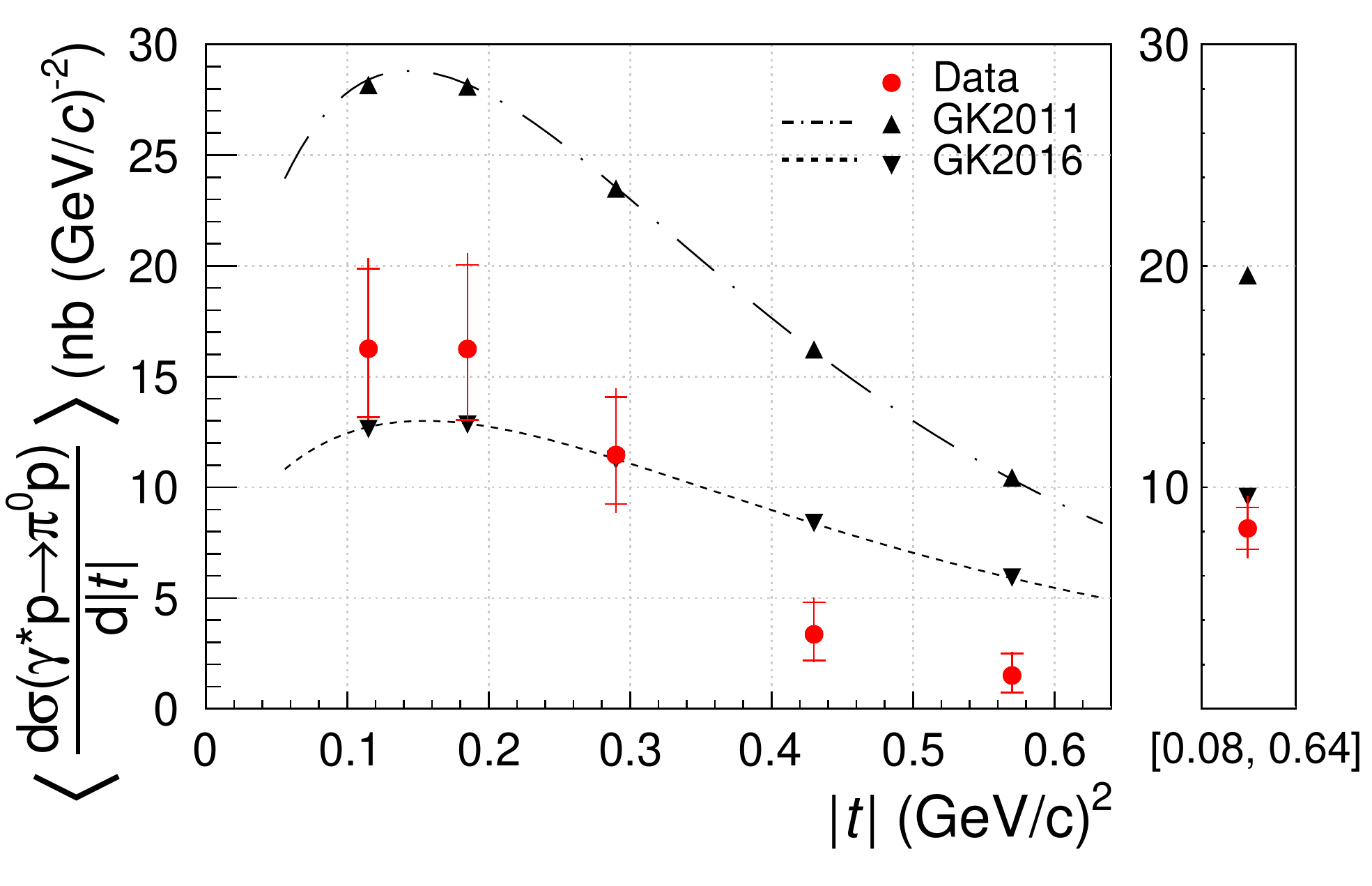}
  \includegraphics[width=\linewidth]{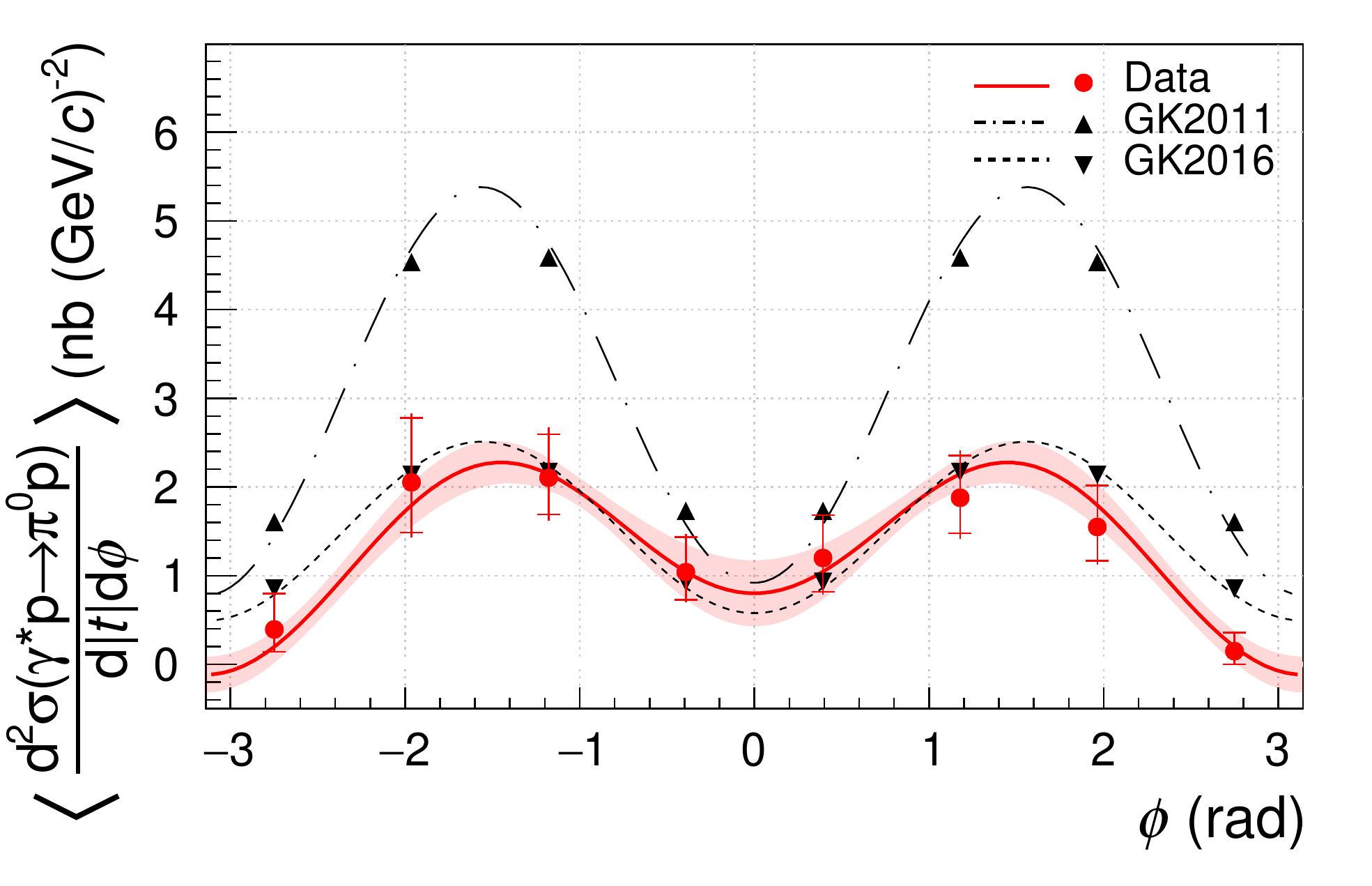}
\end{center} 
\vspace{-0.25cm}
  \caption{\label{fig:x_sec_phi}
    Average value of the differential virtual-photon proton cross section
    $\langle \frac{\di \sigma}{\di |t|} \rangle$ as a function of $|t|$ (top) and
    $\langle \frac{\di^2 \sigma}{\di |t| \di \phi} \rangle$ as a function of $\phi$ (bottom).
    For the top panel the data was integrated over $\phi$, while for the bottom panel it was
    integrated over $|t|$. The result of an integration over $\phi$ and $|t|$ is shown 
    in the  right-most part 
    of the top panel. Inner error bars indicate the statistical uncertainty, outer error
    bars the quadratic sum of statistical and systematic uncertainties. The data is compared
    with two predictions of the  GK model~\cite{GK2011,GKprivate}.
    Radiative corrections are not applied but an estimate is included in the systematic uncertainties.
   }
\end{figure}

For the background corrected final data sample the average kinematics
are 
    $\langle Q^2 \rangle =2.0\; {(\text{GeV}/c)^2}$,
    $\langle \nu \rangle = 12.8\; {\text{GeV}}$, 
    $\langle x_{Bj} \rangle = 0.093  $ and 
    $\langle -t \rangle = 0.256\; {(\text{GeV}/c)^2}  $.
The dependences of the measured cross section on $\abs{t}$ and $\phi$ are shown
in Fig.~\ref{fig:x_sec_phi}, with the numerical values given in Table~\ref{tab:2}. 
The cross section in bins of $|t|$ is shown in the top panel of Fig.~\ref{fig:x_sec_phi}.
It appears to be consistent with an exponential decrease with
increasing $|t|$ for values of $|t|$ larger
than about 0.25 (GeV/$c$)$^2$, while at smaller $|t|$  the $t$-dependence becomes weaker. 
Our result is compared to the predictions of two versions of the Goloskokov-Kroll 
(GK) model~\cite{GK2011,GKprivate}. The results of the GK model shown in this letter
are obtained by integrating over the analysis range in the same way as it is done for the data.
The dashed-dotted curve represents the cross
section from the earlier version~\cite{GK2011} as a function of $|t|$, while the 
upwards pointing triangles correspond to the cross section
averaged over $|t|$ bins of the data. The mean cross sections for the full $t$-range are compared 
in the rightmost part of 
this panel. Analogously, the dotted curve and the downward pointing triangles 
correspond to the later version of the model~\cite{GKprivate}, which was inspired by the results presented in this
Letter. 
We observe that for the earlier version the magnitude of the predicted cross section overshoots 
our measurement by approximately a factor of two.

The cross section as a function of $\phi$ 
for the full measured $t$-range is shown in the bottom panel of Fig.~\ref{fig:x_sec_phi} in eight 
$\phi$ bins of equal width. The full dots 
show the measured cross section for each bin and 
the solid curve represents the fit described below.

In order to extract the different 
contributions to the spin-independent cross 
section, a maximum-likelihood fit is applied to the data according to Eq.~(\ref{eq:new_unpo}).
In the fit, the measured average value of the virtual-photon polarisation parameter is used, $\epsilon = 0.996$.
The $\phi$-integrated cross section determined 
by the fit is obtained as
\begin{equation}
\Big\langle \frac{\di \sigma_{{T}}}{\di |t|} + \epsilon \frac{\di \sigma_{{L}}}{\di |t|} \Big\rangle 
= (8.1 \pm 0.9_{\text{stat}}{}_{- \ 1.0}^{+ \ 1.1}\big\rvert_{\text{sys}})\frac{\text{nb}}{(\text{GeV}/c)^2}.
\end{equation}
The $TT$ and $LT$ interference terms are obtained as
\begin{equation}
\Big\langle \frac{\di \sigma_{{TT}}}{\di |t|} \Big\rangle = (-6.0 \pm 1.3_{\text{stat}}{}_{- \ 0.7}^{+ \ 0.7}\big\rvert_{\text{sys}})\frac{\text{nb}}{(\text{GeV}/c)^2}
\end{equation}
and
\begin{equation}
\Big\langle \frac{\di \sigma_{{LT}}}{\di |t|} \Big\rangle = (1.4 \pm 0.5_{\text{stat}}{}_{- \ 0.2}^{+ \ 0.3}\big\rvert_{\text{sys}})\frac{\text{nb}}{(\text{GeV}/c)^2}.
\end{equation}

We observe a large negative contribution 
by $\sigma_{TT}$ and a smaller positive one 
by $\sigma_{LT}$, which indicates a significant role of transversely polarised photons in exclusive $\pi ^0$ production.

The $\phi$-dependence of the measured cross section is compared to the calculations of the GK model 
in the bottom panel of Fig.~\ref{fig:x_sec_phi}. Apart from the discrepancy in the magnitude of cross sections mentioned before,
here we observe also different
shapes 
for the measurement and the model
predictions, which indicates that the relative contributions of the interference terms
$\sigma_{TT}$ and $\sigma_{LT}$ are different 
when comparing measurement and 
model.

\begin{table}[h!] 
\caption{\label{tab:2}
Numerical values of the average cross sections shown in Fig.~\ref{fig:x_sec_phi}.
}
\footnotesize
\begin{center}
\begin{tabular}{r c  c c}
{\footnotesize lower} & $\Big\langle \frac{\di \sigma}{\di |t| \di \phi} \Big\rangle / \frac{\text{nb}}{(\text{GeV}/c)^2}$ &
{\footnotesize lower}& $\Big\langle \frac{\di \sigma}{\di |t|} \Big\rangle / \frac{\text{nb}}{(\text{GeV}/c)^2}$\\[-4pt]
{\footnotesize $\phi$ bin} & & {\footnotesize $|t|$ bin} & \\
{\footnotesize limit} & & {\footnotesize limit}& \\[2pt]
\hline\\[-6pt]
$-\pi$ & \xsecval{0.4}{0.3}{0.4}{0.1}{0.1} & $0.08$ & \xsecval{16.3}{3.1}{3.6}{2.0}{2.0} \\[4pt]
$\frac{-3\pi}{4}$ & \xsecval{2.1}{0.6}{0.7}{0.2}{0.3} & $0.15$ & \xsecval{16.2}{3.2}{3.8}{1.9}{2.1} \\[4pt]
$\frac{-\pi}{2}$ & \xsecval{2.1}{0.4}{0.5}{0.3}{0.3} & $0.22$ & \xsecval{11.5}{2.2}{2.6}{1.4}{1.5} \\[4pt]
$\frac{-\pi}{4}$ & \xsecval{1.0}{0.3}{0.4}{0.1}{0.2} & $0.36$ & \xsecval{\textcolor{white}{0}3.4}{1.2}{1.4}{0.5}{0.8} \\[4pt]
$0$ & \xsecval{1.2}{0.4}{0.5}{0.2}{0.2} & $0.5$ & \xsecval{\textcolor{white}{0}1.5}{0.8}{1.0}{0.2}{0.4} \\[4pt]
$\frac{\pi}{4}$& \xsecval{1.9}{0.4}{0.5}{0.2}{0.2} &  &  \\[4pt]
$\frac{\pi}{2}$ & \xsecval{1.5}{0.4}{0.5}{0.2}{0.2} &  &  \\[4pt]
$\frac{3\pi}{4}$ & \xsecval{0.1}{0.1}{0.2}{0.0}{0.1} &  &  \\[4pt]
\end{tabular}
\end{center}
\end{table}

{According to Refs.~\cite{GK2011,clas1}, the 
different terms contributing two the cross section for exclusive
pseudoscalar meson production, which appear in Eq.~(\ref{eq:new_unpo}), depend on GPDs $\tilde{H}$, $\tilde{E}$,
$H_T$ and $\overline{E}_T$ = 2$\tilde{H}_T$ + $E_T$. For $\pi^0$ production a large contribution
from transversely polarised virtual photons is expected, which is mainly generated by the chiral-odd
GPD $\overline{E}_T$. It manifests itself in a large contribution from $\sigma _{TT}$ and a dip in the
differential cross section ${\rm d}\sigma /{\rm d} t$ as $|t|$ decreases to zero.
These features are in
qualitative agreement with our results, as also with earlier measurements at different
kinematics~\cite{clas2,clas1,hallA2}. The COMPASS results on exclusive
$\pi^0$ production provide significant constraints on modelling the chiral-odd
GPDs, in particular GPD $\overline{E}_{T}$.}

\section{Summary and conclusion}
  Using exclusive $\pi^0$ muoproduction we have measured the $t$-dependence
  of the virtual-photon proton cross section for hard exclusive $\pi^0$ production 
  at 
    $\langle Q^2 \rangle =2.0\; {(\text{GeV}/c)^2}$
    $\langle \nu \rangle = 12.8\; {\text{GeV}}$, 
    $\langle x_{Bj} \rangle = 0.093$ 
    and 
    $\langle -t \rangle = 0.256\; {(\text{GeV}/c)^2}  $.
  Fitting the azimuthal dependence reveals 
  a large negative contribution 
by $\sigma_{TT}$ and a smaller positive one 
by $\sigma_{LT}$, which indicates a significant role of transversely polarised photons in exclusive $\pi ^0$ production.
  These results provide important input for modelling Generalised Parton Distributions.
  In the context of the phenomenological GK model, the statistically significant $TT$ contribution
  constitutes clear experimental evidence for the existence of the chiral-odd
  GPD $\overline{E}_{T}$.

\section*{Acknowledgements}
We thank Sergey Goloskokov and Peter Kroll for their continuous support with model predictions,
Pierre Guichon for the evaluation of the Bethe-Heitler contribution taking into account
the muon mass. We gratefully acknowledge the support of the CERN
management and staff and the skill and effort of the technicians of our collaborating institutes.  

\section*{References}

\end{document}